\documentclass[reqno]{amsart}

\usepackage{amsmath,amssymb,amsthm,amsfonts,mathrsfs,textcomp,mathtools,accents}
\usepackage{graphicx}
\usepackage{enumerate}
\usepackage[shortlabels]{enumitem}
\setenumerate{label={\upshape(\roman*)},wide=0pt,topsep=0.2cm,labelsep=0.1em,itemsep=0.15cm,leftmargin=*}
\setlist[1]{topsep=0.2cm,itemsep=0.2cm}

\usepackage[headings]{fullpage}
\usepackage{hyperref}
\hypersetup{
  colorlinks=false
}
\usepackage{bm}
\usepackage{natbib}

\DeclareFontFamily{U}{mathx}{\hyphenchar\font45}
\DeclareFontShape{U}{mathx}{m}{n}{<-> mathx10}{}
\DeclareSymbolFont{mathx}{U}{mathx}{m}{n}
\DeclareMathAccent{\widebar}{0}{mathx}{"73}

\usepackage[figuresright]{rotating}
\usepackage{tikz}
\usetikzlibrary{arrows,calc,positioning}

\usepackage{booktabs}

\newtheorem{theorem}{Theorem}[section]

\theoremstyle{definition}
\newtheorem{definition}[theorem]{Definition}

\theoremstyle{remark}
\newtheorem{remark}[theorem]{Remark}

\numberwithin{equation}{section}

\newcommand{\R}{\mathbb R}

\newcommand{\clomon}{{\sc clomon}}

\usepackage{xcolor}
\definecolor{dgreen}{RGB}{0,160,0}    
\definecolor{orange}{RGB}{251,111,66} 
\definecolor{lblue}{RGB}{8,180,238}   
\definecolor{DBlu}{rgb}{.1,.1,.7}     

\begin{document}

\title[The Manifold Of Variations: impact location of short-term
impactors]{The Manifold Of Variations:\\impact location of short-term
  impactors}%

\author{Alessio Del Vigna}%
\address{Space Dynamics Services s.r.l., via Mario Giuntini,
  Navacchio di Cascina, Pisa, Italy}
\address{Dipartimento di Matematica, Universit\`a di Pisa, Largo Bruno
  Pontecorvo 5, 56127 Pisa, Italy} \email{delvigna@mail.dm.unipi.it}

\author{Linda Dimare}
\address{Space Dynamics Services s.r.l., via Mario Giuntini,
  Navacchio di Cascina, Pisa, Italy}

\author{Davide Bracali Cioci}
\address{Space Dynamics Services s.r.l., via Mario Giuntini,
  Navacchio di Cascina, Pisa, Italy}

\begin{abstract}
    The interest in the problem of small asteroids observed shortly
    before a deep close approach or an impact with the Earth has grown
    a lot in recent years. Since the observational dataset of such
    objects is very limited, they deserve dedicated orbit
    determination and hazard assessment methods. The currently
    available systems are based on the systematic ranging, a technique
    providing a 2-dimensional manifold of orbits compatible with the
    observations, the so-called Manifold Of Variations. In this paper
    we first review the Manifold Of Variations method, to then show
    how this set of virtual asteroids can be used to predict the
    impact location of short-term impactors, and compare the results
    with those of already existent methods.
\end{abstract}

\maketitle

\section{Introduction}
\label{sec:intro}

Astronomers observe the sky every night to search for new asteroids or
for already known objects. The Minor Planet Center (MPC) collects the
observations coming from all over the world and then tries to compute
orbits and to determine the nature of the observed objects. New
discoveries which could be Near-Earth Asteroids (NEAs) are posted in
the NEO Confirmation Page
(NEOCP\footnote{\url{https://minorplanetcenter.net/iau/NEO/toconfirm_tabular.html}}),
which thus contains observational data of unconfirmed objects. They
could be real asteroids as well as non-real objects, and cannot be
officially designated until additional observations are enough to
compute a reliable orbit and confirm the discovery.

Some asteroids with an Earth-crossing orbit may impact our planet and
the goal of impact monitoring is to identify potentially hazardous
cases and solicit follow-up observations. Two automated and
independent systems continually scan the catalogue of known NEAs with
this purpose, namely
\clomon-2\footnote{\url{https://newton.spacedys.com/neodys/index.php?pc=0}}
and Sentry\footnote{\url{http://cneos.jpl.nasa.gov/sentry/}}, which
are respectively operational at the University of Pisa/SpaceDyS and at
the NASA JPL. Nevertheless, also objects waiting for confirmation in
the NEOCP could be on a collision trajectory with the Earth, sometimes
with the impact occurring just a few hours after the discovery. This
is exactly what happened for asteroids 2008~TC$_3$, 2014~AA, 2018~LA,
and 2019~MO, all discovered less than one day prior to striking the
Earth and prior to being officially designated by the MPC. Thus being
able to perform a reliable hazard assessment also in these cases is a
fundamental issue, but needs dedicated techniques due to the very
different nature of the problem. Indeed when an asteroid is first
observed, the available astrometric observations are few and cover a
short time interval. This amount of information is usually not enough
to allow the determination of a well-constrained orbit and in fact the
orbit determination process shows some kind of degeneracies. The few
observations constrain the position and motion of the object on the
celestial sphere, but leave practically unknown the topocentric range
and range-rate. As a consequence the set of orbits compatible with the
observations forms a region in the orbital elements space which has a
two-dimensional structure, so that every one-dimensional
representation of the region such as the Line Of Variations (LOV,
\cite{milani:multsol}) would be unreliable.

Two systems are now publicly operational and dedicated to the orbit
determination and hazard assessment of unconfirmed objects: Scout at
the NASA JPL \citep{farnocchia2015} and NEOScan at SpaceDyS
\citep{spoto:immimp}. They are both based on the systematic ranging,
an orbit determination method which explores a subset of admissible
values for the range and
range-rate. NEOScan\footnote{\url{https://newton.spacedys.com/neodys2/NEOScan/}}
makes use of the Admissible Region (AR, \cite{milani2004AR}) as a
starting point to explore the range and range-rate space. Then the
short arc orbit determination process ends with the computation of the
Manifold Of Variations (MOV, \cite{tommei:phd}), a 2-dimensional
compact manifold of orbits parameterized over the AR. A finite
sampling of the MOV is thus a suitable representation of the
confidence region, because it accounts for its two-dimensional
structure, and thus the resulting set of virtual asteroids can be used
for the short-term hazard assessment of such objects
\citep{spoto:immimp,delvigna:mov1}. A part of this
activity is the prediction of the impact location of a potential
impactor, especially when the associated impact probability is high. A
method to predict the impact corridor of an asteroid has been
developed in \cite{dimare:imp_corr}, by which the impact region is
given by semilinear boundaries on the impact surface at a given
altitude above the Earth and corresponding to different confidence
levels. The algorithm is conceived to be a continuation of the impact
monitoring algorithm at the basis of the \clomon-2 system, since the
semilinear method requires a nominal orbit obtained by full
differential corrections and an impacting orbit\footnote{More
  precisely a representative of the virtual impactor, a connected
  subset of the asteroid confidence region made up of impacting
  orbits.}, as provided by the LOV method \citep{milani:clomon2}. When
the observational arc is very short and the object is waiting for
confirmation in the NEOCP, the semilinear method could become
inapplicable because very often a reliable nominal orbit simply does
not exist. In this paper we propose a method which uses the MOV to
predict the impact location of an imminent impactor, and we then test
this technique using the data available for the four impacted
asteroids so far, namely 2008~TC$_3$, 2014~AA, 2018~LA, and 2019~MO.

In Section~\ref{sec:armov} we give a brief recap about the AR and the
MOV method. In Section~\ref{sec:imploc} we introduce the impact
surface at a given height over the Earth and the impact map, to then
describe how to exploit the MOV sampling for impact location
predictions. Section~\ref{sec:test} contains the results of our method
applied to the impacted asteroids 2008~TC$_3$, 2014~AA, 2018~LA, and
2019~MO. When possible, we also compare the impact regions with those
computed with the semilinear method \citep{dimare:imp_corr} and with a
Monte Carlo simulation. Lastly, Section~\ref{sec:conc} is dedicated to
the conclusions.

\section{The Manifold Of Variations method}
\label{sec:armov}

Suppose we have a short arc of observations, typically $3$ to $5$
observations over a time span shorter than one hour. In most cases the
arc is too short to allow the determination of a full orbit and it is
called a Too Short Arc (TSA, \cite{milani2007}). When in presence of a
TSA, either preliminary orbit methods fail or the differential
corrections do not converge to a nominal orbit. Nevertheless, as
anticipated in the introduction, the little information contained in
the arc can be summarised in the \emph{attributable}, the vector
\[
    \mathcal{A} \coloneqq (\alpha, \delta, \dot{\alpha}, \dot{\delta})
    \in \mathbb{S}^1 \times \left(-\tfrac{\pi}2, \tfrac{\pi}2\right)
    \times \R^2
\]
formed by the angular position and velocity of the object at the mean
observational time. Note that if there is at least one measurement of
the apparent magnitude, the attributable could also contain an average
value for this quantity. The topocentric range $\rho$ and range-rate
$\dot\rho$ are thus not known, otherwise we would have had a full
orbit. The AR comes into play here, to provide a set of possible
values of $(\rho,\dot\rho)$ by imposing general conditions on the
object's orbit. It can be shown that the AR is a compact subset of
$\R_{\geq 0}\times \R$, which can have at most two connected
components. For the precise definition of the AR and the proof of its
properties the reader can refer to \cite{milani2004AR}, here we limit
ourselves to the general idea. We essentially impose that the object
is a Solar System body and that it is sufficiently large to be source
of meteorites. Indeed short-term impactors are usually very small
asteroids, with a few meters in diameter, thus the main interest in
such objects is not for planetary defence, but rather to observe them
during the last part of their impact trajectory and possibly to
recover meteorites on ground, as it happened for asteroid 2008~TC$_3$.

\begin{figure}[h]
    \includegraphics[scale=0.4]{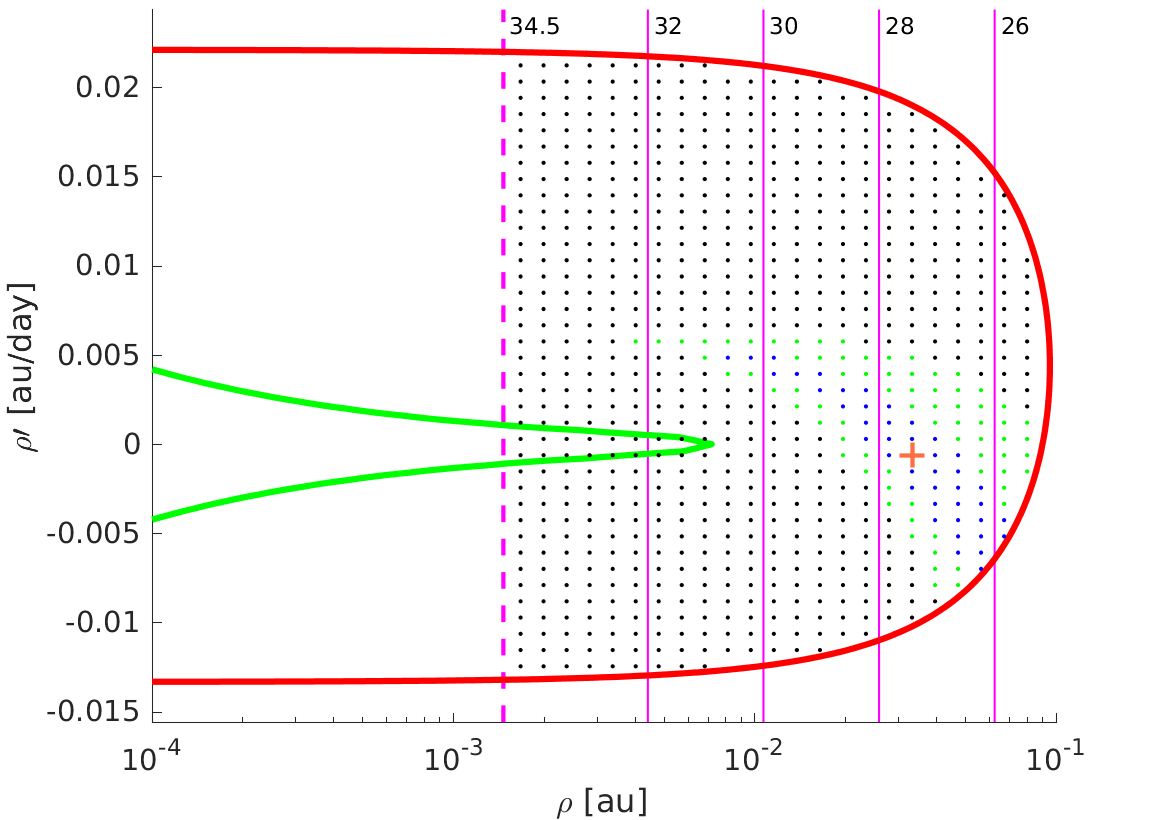}
    \includegraphics[scale=0.4]{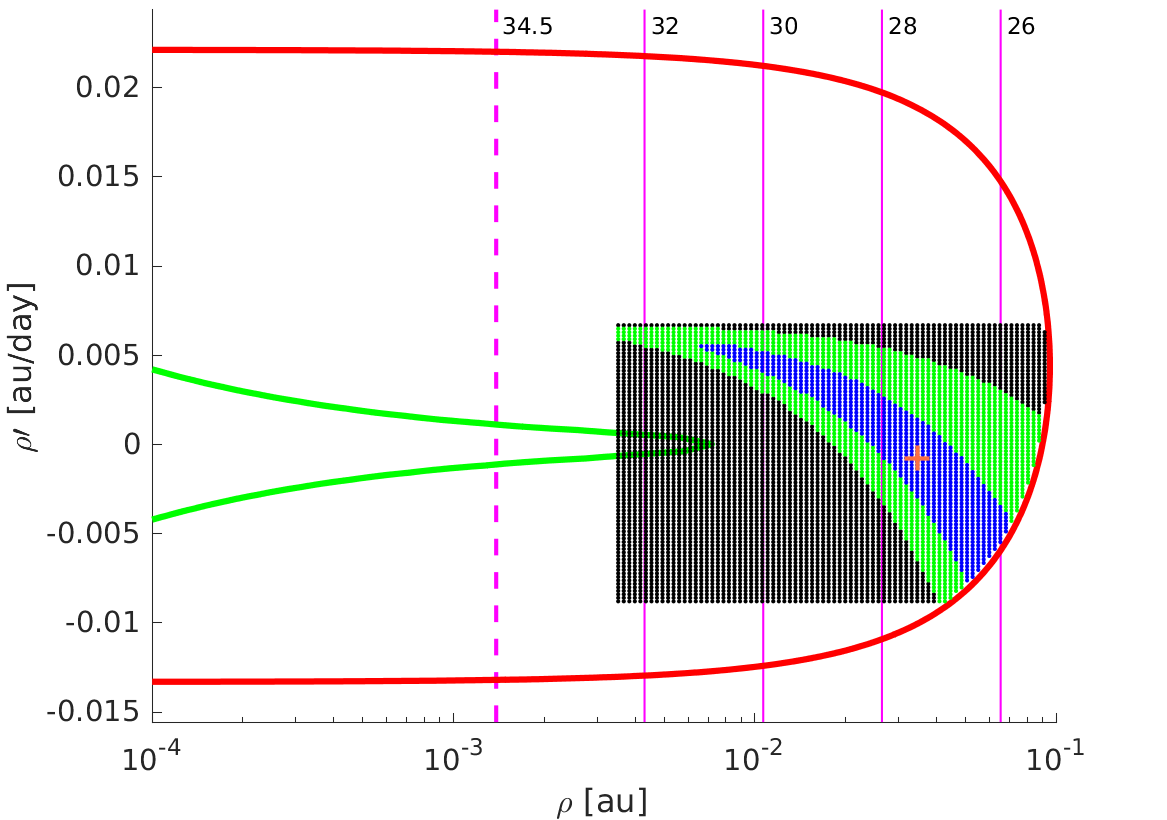}
    \caption{Admissible Region sampling with the rectangular
      grids. \emph{Left}. First step, with the rectangular grid
      covering the entire AR. \emph{Right}. Second grid, covering the
      subregion of good orbits identified in the first step. In both
      plots the sample points are marked in blue when $\chi\leq 2$ and
      in green when $2<\chi<5$. The orange cross marks the orbit with
      the minimum $\chi$ value.}
    \label{fig:grid}
\end{figure}

\begin{figure}[h]
    \centering \includegraphics[scale=0.4]{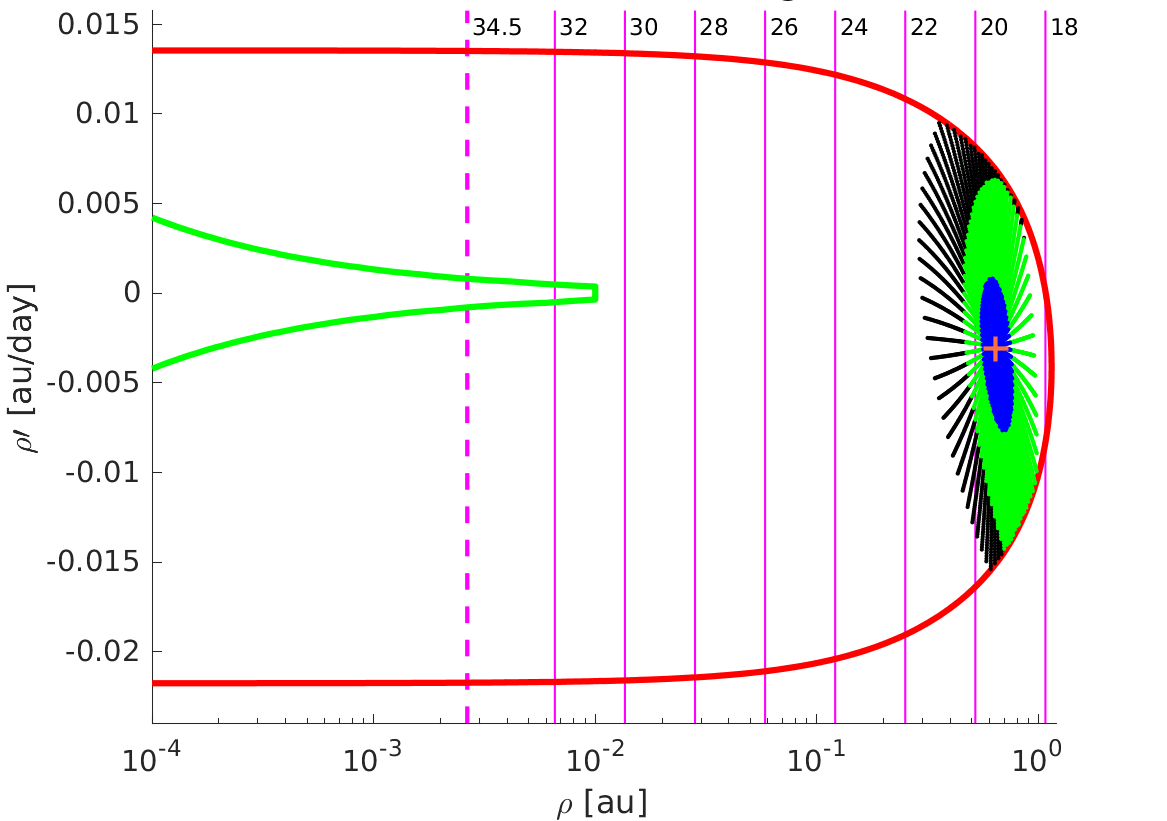}
    \caption{Sampling of the AR with a cobweb around the nominal
      solution . The sample points are marked in blue when
      $\chi\leq 2$ and in green when $2<\chi<5$. The orange cross
      marks the orbit with the minimum $\chi$ value.}
    \label{fig:cobweb}
\end{figure}

The AR is sampled by a finite number of points and with different
techniques. In case of a TSA we explore the whole AR with rectangular
grids: first we cover the entire region and compute the corresponding
sample of orbits, then we identify the subregion corresponding to the
best-fitting orbits with respect to the observations, and lastly we
cover this subregion with a second grid. The two-step procedure is
shown in Figure~\ref{fig:grid}. Even if this is not the most common
case, it may happen that the short arc of observations allows the
computation of a preliminary orbit and even of a full orbit. In this
case we consider the nominal solution as \emph{reliable} only if the
arc curvature is significant \citep{milani2007}. Given its importance
for orbit determination purposes, we give more weight to the geodesic
curvature with respect to the along-track acceleration, imposing that
the signal-to-noise ratio of the geodesic curvature is greater than
$3$. In this case we switch to a different sampling method to exploit
the additional strong information coming with the reliable nominal
orbit. Indeed we consider the marginal covariance of the orbit in the
range and range-rate space, whose probability density function has
level curves which are concentric ellipses around the nominal range
and range-rate. We select a maximum confidence threshold and sample
the AR by following these ellipses up to this confidence level. This
cobweb technique is shown in Figure~\ref{fig:cobweb}. Full details
about the sampling of the AR in the various cases can be found in
\cite{spoto:immimp}.

Now we describe how to obtain a sampling of orbits, namely the MOV,
once a sampling of the AR is given. We start with some notation. The
target function is
\[
    Q(\mathbf{x}) \coloneqq \frac{1}{m} \bm{\xi}(\mathbf x)^\top
    \bm{\xi}(\mathbf x),
\]
where $\mathbf{x}$ are the orbital elements, $m$ is the number of
observations used for the fit, and $\bm{\xi}$ is the vector of the
normalised observed-computed debiased astrometric residuals. Let
$\mathcal{A}_0$ be the attributable computed from the arc of
observations. The AR sampling is a set $K$ of admissible values of the
range and range-rate. For each $\bm\rho_0 = (\rho_0,\dot\rho_0)\in K$
we consider the full orbit $(\mathcal{A}_0,\rho_0,\dot\rho_0)$ and fit
only the first four components, that is the attributable part, with a
constrained differential corrections procedure.

\begin{definition}
    Let $K$ be a subset of the AR. The \emph{Manifold Of Variations}
    is the set $\mathcal M$ of points
    $(\mathcal A^*(\bm\rho_0),\bm\rho_0)$ such that $\bm\rho_0\in K$
    and $\mathcal A^*(\bm\rho_0)$ is the local minimum of the target
    function
    $\left.Q(\mathcal{A},\bm\rho)\right|_{\bm\rho=\bm\rho_0}$, when it
    exists.
\end{definition}

We call $K'\subseteq K$ the set of values of $(\rho,\dot\rho)$ such
that the constrained differential corrections converge, giving an
orbit on $\mathcal{M}$. To estimate the goodness of the fit to the
observations, for each orbit $\mathbf{x}\in \mathcal{M}$ we compute
\[
    \chi(\mathbf x) \coloneqq \sqrt{m(Q(\mathbf x)- Q^*)},
\]
where $Q^*$ is the minimum value of the target function over $K'$. We
always consider MOV orbits having $\chi<5$, thus corresponding at most
to the $5\sigma$ confidence level. For hazard assessment, this
guarantees to find impact possibilities with a probability $>10^{-3}$,
the so-called completeness level of the impact monitoring system
\citep{delvigna:compl_IM}.

Hereafter we summarise the main steps of the Manifold Of Variations
method, as it is implemented in the software system NEOScan.
\begin{enumerate}
  \item The MPC NEO Confirmation Page is scanned every two minutes to
    look for new objects or for new observations to add to previous
    detections.
  \item Computation of the attributable and sampling of the Admissible
    Region with rectangular grids or with the cobweb techinque,
    depending on the existence of a reliable nominal solution.
  \item Computation of the Manifold Of Variations by constrained
    differential corrections, obtaining a set of virtual asteroids
    with a two-dimensional structure.
  \item Propagation of the virtual asteroids for 30 days and
    projection of the propagated MOV sampling on the Modified Target
    Plane\footnote{For an asteroid having a hyperbolic close approach
      with a planet, the Modified Target Plane is the plane passing
      through the centre of the planet and orthogonal to the velocity
      of the body at the time of closest approach.}.
  \item Searching for impacting solutions and, if any, computation of
    the impact probability. The computation of this quantity is done
    with a propagation of the probability density function from the
    normalised residuals space to the sampling space and by
    integrating the resulting density over the set of impacting sample
    points \citep{delvigna:mov1}.
\end{enumerate}


\section{Impact location prediction}
\label{sec:imploc}

Before describing how to exploit the MOV sampling for impact
predictions, we emphasise a key difference in the treatment of
short-term impactors with respect to the long-term hazard monitoring
of known asteroids. The MOV method does not use interpolation of the
sample to find impacting orbits as it is done for the LOV method
\citep{milani:clomon2}, but just considers the impacting orbits of the
sampling. For instance, if none of the MOV sample points impacts the
Earth, we assign a null impact probability. Indeed we adopt a dense
sampling, thus if a set of impacting orbits is not detected by the
sampling it has a too small probability to be interesting. This is a
crucial difference with respect to the LOV method at the basis of
long-term impact monitoring. Many of the NEAs in the catalogue are big
objects and thus also impact events with probability of few parts per
million are worth detecting. Since nearby LOV orbits are separated by
the chaotic dynamics introduced by close approaches, the prediction of
impacts with such low probabilities, especially if far in time with
respect to the time of the observations, would require an extremely
high number of sample points along the LOV. Hence if we limited the
analysis to the sample points only, the computations would be too
heavy. As a consequence, interpolation of the LOV sampling is
essential\footnote{Actually in some extremely non-linear cases this is
  not enough, and LOV sampling densification techniques have been
  developed to overcome the problem and to guarantee that the preset
  completeness level is reached \citep{delvigna:dens}.}, but with the
MOV method we are allowed to consider just the impacting orbits of the
sampling.

On the basis of the above comment, the short-term hazard analysis
performed with the MOV method ends with the identification of a subset
$\mathcal{V}$ of impacting orbits among the MOV sampling (see step (v)
at the end of the previous section). The idea of our location
prediction method is actually very simple: once the set $\mathcal{V}$
is given, we just propagate the orbits of $\mathcal{V}$ until the
impact and compute the geodetic coordinates of the impacting
points.

For such predictions we consider the WGS 84 model \citep{wgs84}, for
which the Earth surface is approximated by a geocentric oblate
ellipsoid with semimajor axis equal to $6378.137$~km and flattening
parameter $f$ defined by $1/f=298.257223563$. The eccentricity $e$ of
the ellipsoid can be computed as $e^2=f(2-f)$.

\begin{definition}
    Let $h\geq 0$. The \emph{impact surface} $S_h$ at altitude $h$
    above ground is the set of points at altitude $h$ above the WGS 84
    Earth ellipsoid.
\end{definition}

\noindent The impact region will be a subset of the impact surface
$S_h$ for a given value of the altitude $h$ and points on $S_h$ are
given by means of the geodetic coordinates. To compute this region we
use the impact map introduced in \cite{dimare:imp_corr}, that is
\[
    F^h: \mathcal{V} \subseteq \mathcal{M} \rightarrow S_h.
\]
This map takes an impacting orbit $\mathbf{x}\in \mathcal{V}$,
computes the time $t(\mathbf{x})$ at which the orbit reaches the
surface $S_h$, applies the integral flow of the system to propagate
the orbit to the time $t(\mathbf{x})$, and lastly converts the state
vector into the geodetic coordinates on $S_h$. Thus our \emph{MOV
  impact region} is the set $F^h(\mathcal{V})\subseteq S_h$, that is
the set of impacting MOV orbits propagated to the surface $S_h$.

\subsection{Semilinear boundaries for the impact corridor prediction}

We briefly recap the semilinear method applied to the problem of the
impact location prediction, described in \cite{dimare:imp_corr}.

Let $\mathcal{X}$ be the orbital elements space, let
$\mathcal{Y}\subseteq \R^2$ be the target space, and let
$F:W\rightarrow \mathcal{Y}$ be a differentiable function defined on
an open subset $W\subseteq\mathcal{X}$, to be thought of as the
prediction function. The dimension $N$ of $\mathcal{X}$ is $6$ if we
consider the six orbital elements, or is $>6$ if some dynamical
parameter is included\footnote{A typical situation is $N=7$ and
  consists in the inclusion of the Yarkovsky effect in the dynamical
  model. The relevance of this non-gravitational perturbation for the
  impact monitoring of some NEAs is well-known. Recent examples are
  presented in \cite{delvigna:410777} for asteroid (410777)~2009~FD
  and in \cite{tholen:99942} for asteroid Apophis, in both cases
  succeeding in ruling out the most relevant virtual
  impactors.}. Furthermore we consider a nominal orbit
$\mathbf{x}^*\in \mathcal{X}$ provided with its $N\times N$ covariance
matrix $\Gamma_X$. The uncertainty of $\mathbf{x}^*$ can be described
through the confidence region
\[
    Z^X(\sigma) \coloneqq \left\{\mathbf{x}\in \mathcal{X} \,:\,
      Q(\mathbf{x})-Q(\mathbf{x}^*) \leq \frac{\sigma^2}{m}\right\},
\]
where $\sigma>0$ is the confidence parameter. The prediction function
$F$ maps the orbital elements space onto the target space, and thus
maps the confidence region $Z^X(\sigma)$ to
$Z^Y(\sigma)\coloneqq F(Z^X(\sigma))$. The semilinear method provides
an approximation of the boundary $\partial Z^Y(\sigma)$ of the
non-linear prediction region \citep{milani:ident2}.

To explain the construction of the semilinear boundaries we start with
the notion of linear prediction. The inverse of $\Gamma_X$, that is
$C_X=\Gamma_X^{-1}$, is the normal matrix and defines the linear
confidence ellipsoid
\[
    Z_{lin}^X(\sigma)\coloneqq \left\{\mathbf{x} \in \mathcal{X} \,:\,
      (\mathbf{x}-\mathbf{x}^*)\cdot C_X (\mathbf{x}-\mathbf{x}^*)
      \leq \sigma^2 \right\}.
\]
The differential $DF_{\mathbf{x}^*}$, assumed to be full rank, maps
$Z_{lin}^X(\sigma)$ to the ellipse
\[
    Z_{lin}^Y(\sigma) \coloneqq \left\{\mathbf{y}\in\mathcal{Y} \,:\,
      (\mathbf{y}-\mathbf{y}^*)\cdot C_Y (\mathbf{y}-\mathbf{y}^*)\leq
      \sigma^2\right\},
\]
where $\mathbf{y}^*=F(\mathbf{x}^*)$ is the nominal prediction,
$C_Y=\Gamma_Y^{-1}$, and
$\Gamma_Y = DF_{\mathbf{x}^*} \Gamma_X DF_{\mathbf{x}^*}^\top$
according to the covariance propagation law. When $F$ is strongly
non-linear, this linear prediction region is a poor approximation of
$Z^Y(\sigma)$ and here the semilinear method comes into play. The
boundary ellipse $\partial Z_{lin}^Y(\sigma)$ is the image by the
linear map $DF_{\mathbf{x}^*}$ of an ellipse
$\mathcal{E}_X(\sigma)\subseteq \mathcal{X}$ which lies on the surface
$\partial Z_{lin}^X(\sigma)$. The \emph{semilinear boundary}
$K(\sigma)$ is the non-linear image in the target space of the ellipse
$\mathcal{E}_X(\sigma)$, that is
\[
    K(\sigma)\coloneqq F(\mathcal{E}_X(\sigma)).
\]
If the curve $K(\sigma)$ is simple (no self-intersections) and closed,
then by the Jordan curve theorem it is the boundary of a connected
subset $Z(\sigma)\subseteq \mathcal{Y}$, which we use as an
approximation of $Z^Y(\sigma)$.

\begin{remark}\label{rem}
    The differential $DF_{\mathbf{x}_0}$ is clearly not
    injective. More precisely, each point of $\mathcal{Y}$ has a
    $(N-2)$-dimensional fiber in $\mathcal{X}$, and thus in principle
    the selection of the ellipse $\mathcal{E}_X(\sigma)$ could be done
    in infinitely many ways. The choice foreseen in the semilinear
    method consists in selecting as $\mathcal{E}_X(\sigma)$ the
    ellipse resulting from the intersection between a suitable
    regression subspace in $\mathcal{X}$ and the confidence ellipsoid
    $Z_{lin}^X(\sigma)$. See \cite{milani:ident2} for all the details.
\end{remark}

The application of this method to the impact corridor problem is
described in \cite{dimare:imp_corr} and works as follows. We have a
nominal orbit $\mathbf{x}_0$ of an asteroid with a non-negligible
chance of impacting the Earth, and thus a virtual impactor
representative $\mathbf{x}_{imp}$ as provided by the LOV method. In
this problem the prediction map is the impact map
\[
    F^h: W \rightarrow S_h
\]
defined as above, where $W\subseteq \mathcal{X}$ is an open
neighbourhood of $\mathbf{x}_{imp}$. The semilinear method can now be
applied, with the following adaptation. For the linear prediction on
the impact surface and the selection of the ellipse
$\mathcal{E}_X(\sigma)$ we use the differential
$(DF^h)_{\mathbf{x}_{imp}}$ of the impact map at the virtual impactor
representative orbit $\mathbf{x}_{imp}$. The result of this method are
curves on the surface $S_h$ corresponding to different values of
$\sigma$. Note that these curves need not be closed in this context,
because in general not all the orbits on $\mathcal{E}_X(\sigma)$
impact.


\section{Numerical tests}
\label{sec:test}

We test the MOV method for the impact location prediction of four
asteroids, namely 2008~TC$_3$, 2014~AA, 2018~LA and 2019~MO. When we
have a reliable nominal solution we also compare these results with
those of the semilinear method and, additionally, to an observational
Monte Carlo simulation. The latter amounts to adding noise to each
observation to then compute a new orbit based upon the revised
observations and use this orbit as a virtual asteroid. When the
observed arc is short the uncertainty in the orbit determination is
typically so large that the true uncertainty is not represented by the
confidence ellipsoid. In this case the observational Monte Carlo is
the best approach to follow, definitely better than an orbital Monte
Carlo, which works well when the orbital uncertainty is fairly good.

\subsection{Asteroid 2008~TC$_3$}
The small asteroid 2008~TC$_3$ was discovered by Richard A. Kowalski
at the Catalina Sky Survey on October 6, 2008. The preliminary orbit
computations done at the MPC immediately revealed that the object was
going to impact the Earth within 21 hours. Thus the astronomical
community made a great effort to observe it and the currently
available dataset contains nearly 900 observations, though not all of
them are of high quality, and need to be properly treated for a
precise estimate of the trajectory of 2008~TC$_3$
\citep{farnocchia2008}.

\begin{figure}[h]
    \includegraphics[scale=0.62,trim={0 0 0 0.5cm},clip]{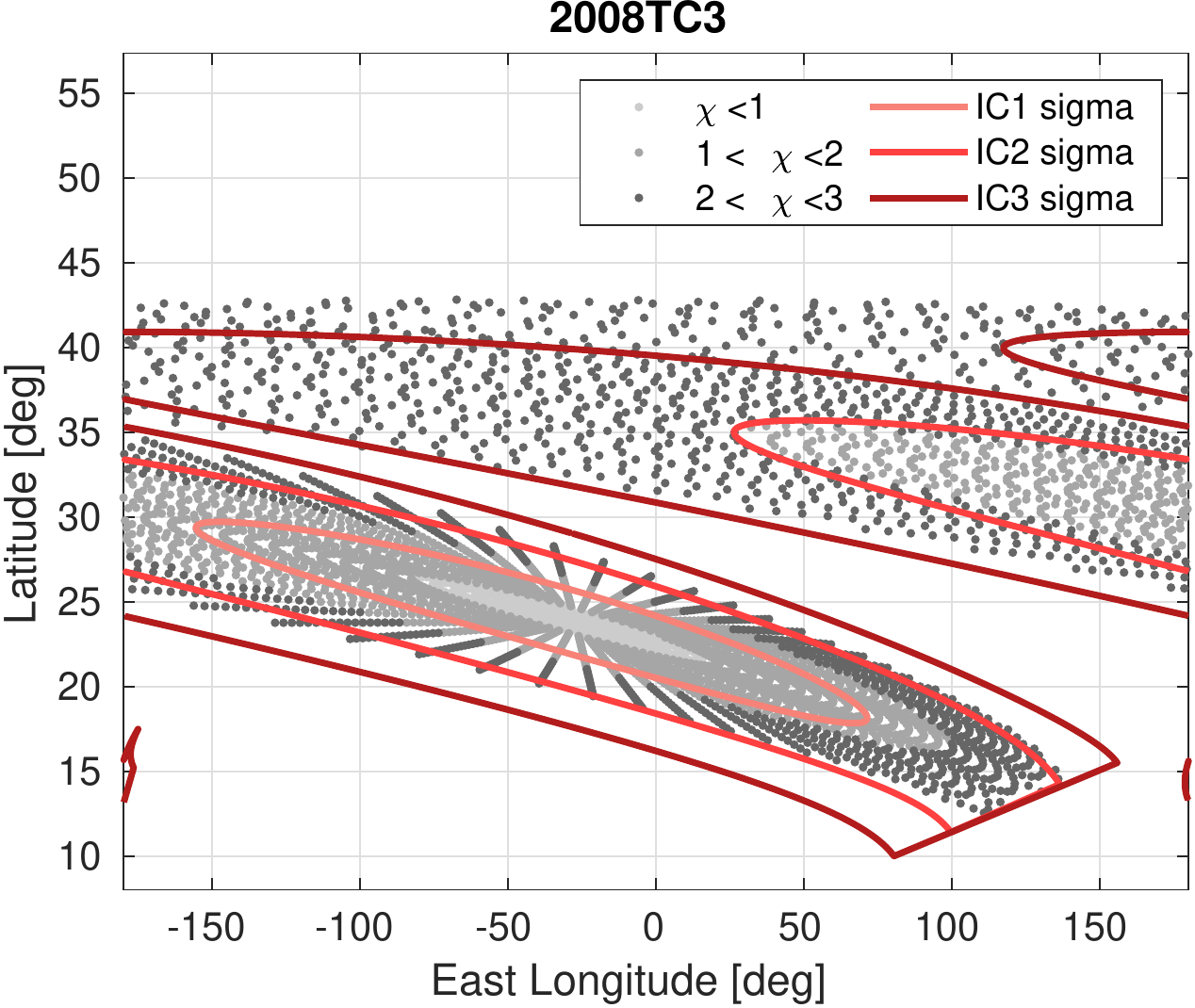}
    \hfill
    \includegraphics[scale=0.62,trim={0 0 0 0.5cm},clip]{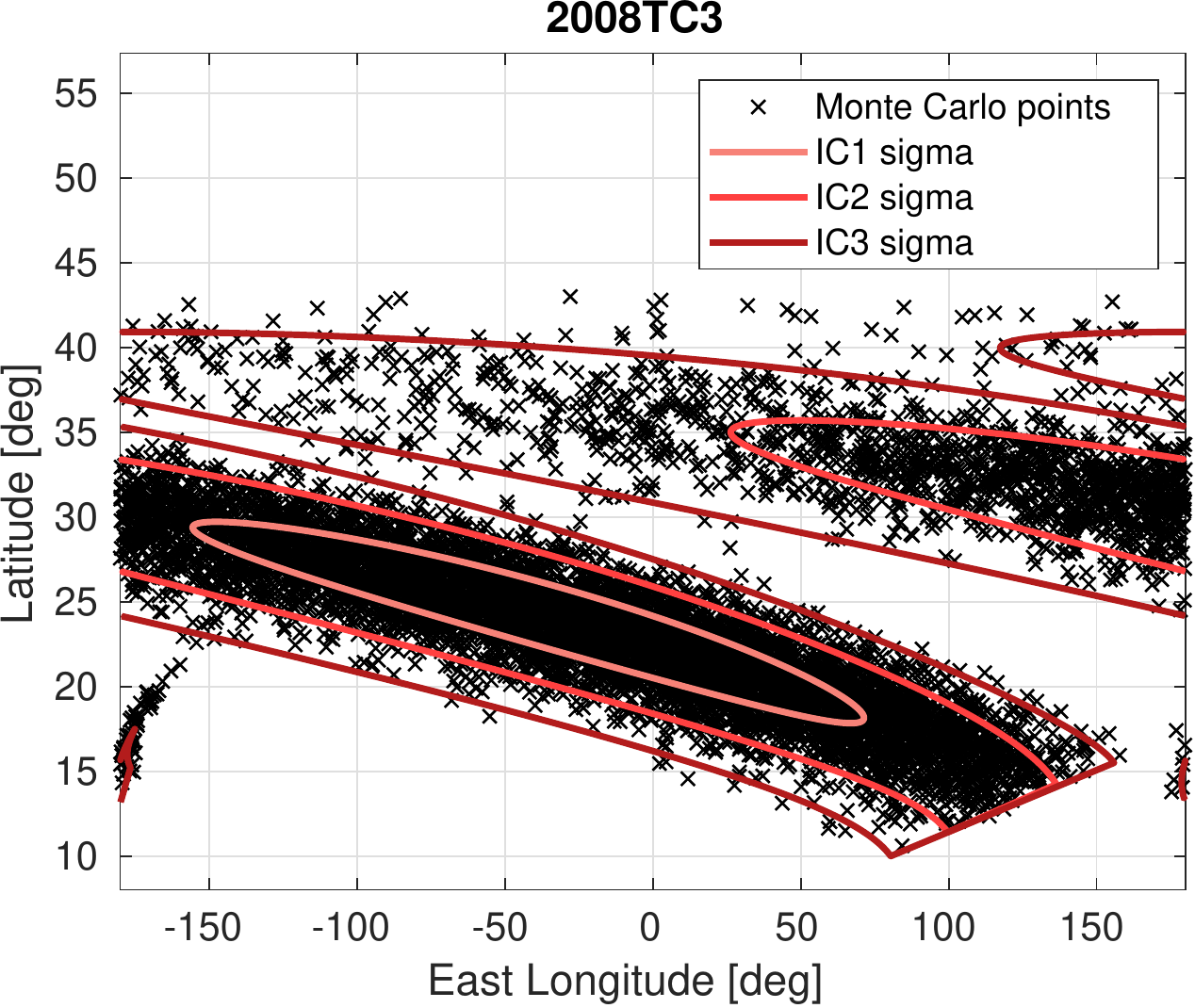}
    \caption{Impact location prediction on ground for asteroid
      2008~TC$_3$. {\em Left.} Comparison between the impact region
      computed with the MOV orbits having $\chi<3$ and the semilinear
      boundaries corresponding to the confidence levels $1$, $2$, and
      $3$. {\em Right.} Comparison between a Monte Carlo run and the
      same semilinear boundaries.}
    \label{fig:2008TC3}
\end{figure}

When asteroid 2008~TC$_3$ was recognised to be an impactor it only had
few observations and for our analysis we consider its first seven
observations. They are enough to compute a reliable nominal solution,
so that our algorithm samples the AR with the cobweb technique. The
application of the MOV method results in an impact probability of
99.7\%, which means that the vast majority of the MOV orbits impacts
the Earth. We then propagate the MOV sampling to the Earth surface,
{\em i.e.} we set $h=0$, and obtain the result shown in
Figure~\ref{fig:2008TC3}. Since the impact region is very extended, we
limit our analysis to the MOV orbits with $\chi<3$. In the left figure
we show the geodetic coordinates of the impacting points, with
different shades of gray depending on the $\chi$ value. In the right
figure we show the results of a Monte Carlo simulation with 10,000
sample points, to compare with the MOV impact region. Moreover, thanks
to the existence of a nominal orbit we can also apply the semilinear
method, thus in both plots of Figure~\ref{fig:2008TC3} we also draw the
semilinear boundaries on ground corresponding to the confidence levels
$1$, $2$, and $3$. The result is a strong agreement of the three
methods.

We notice that the semilinear boundaries enclose a region which is
larger than that obtained with the MOV method. This happens also in
most of the other examples which we present in the next sections. We
can explain this behaviour as follows. Recall that the basic idea of
the semilinear method is to select a curve in the orbital elements
space to propagate non-linearly to approximate the boundary of the
non-linear prediction region. The choice of this curve is made by
considering the boundary of the marginal uncertainty on a suitable
space, which is equivalent to considering the corresponding regression
subspace (see Remark~\ref{rem}). The marginal covariance is the
largest projected uncertainty on a given space thus, so to say, is the
most conservative choice.

\subsection{Asteroid 2014~AA}
Asteroid 2014~AA was discovered by R. Kowalski at the Catalina Sky
Survey on January 1, 2014 at 06:18 UTC. Similarly to 2008~TC$_3$, also
2014~AA impacted the Earth just 21 hours after its first
detection. But very differently from 2008~TC$_3$, due to the
particular night in which 2014~AA was spotted, it was not recognised
to be an impactor. As a consequence the astrometric dataset is very
limited, containing just 7 observations.

\begin{figure}[h!]
    \includegraphics[scale=0.62,trim={0 0 0 0.5cm},clip]{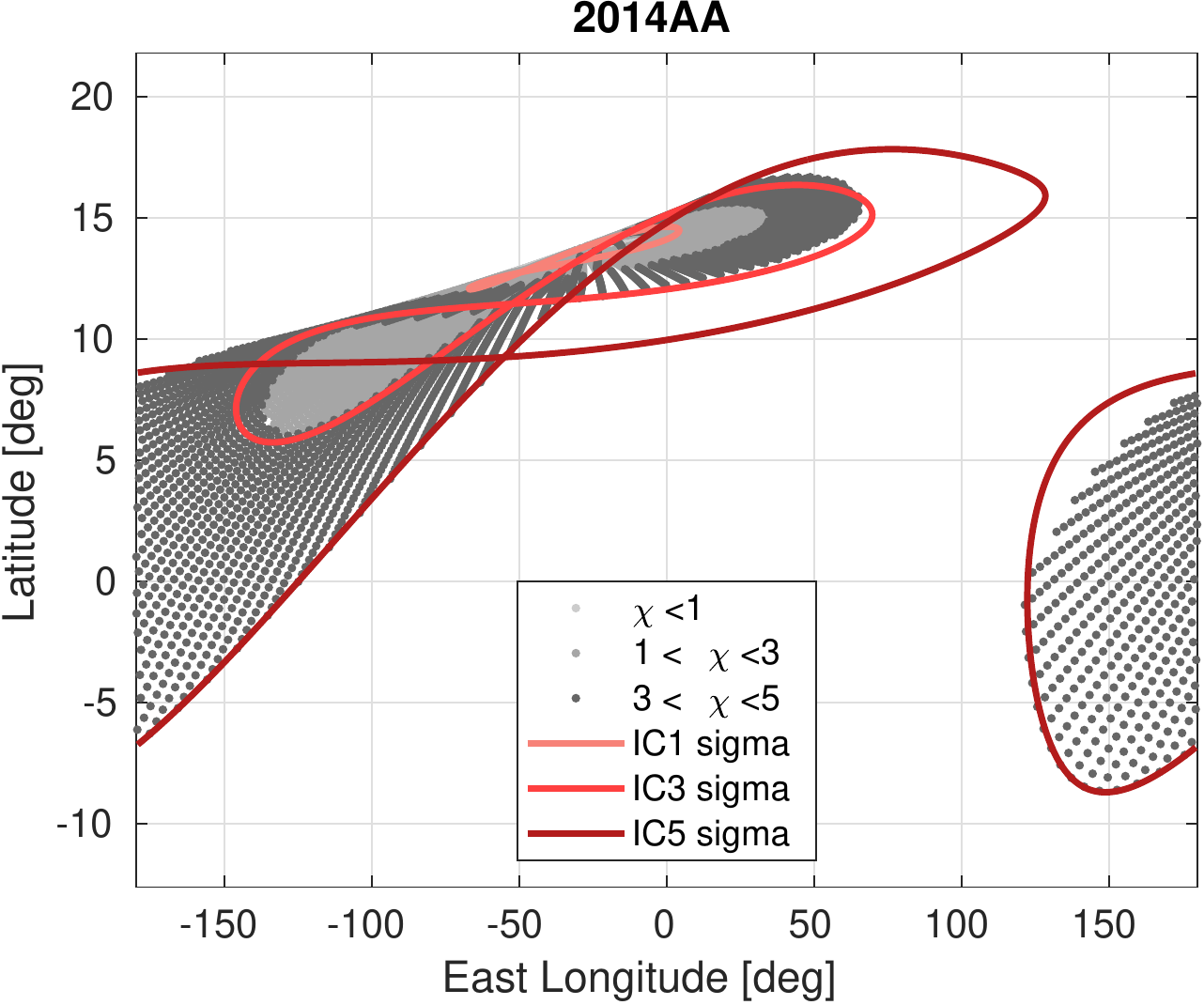}
     \hfill
     \includegraphics[scale=0.454,trim={0 0 0 0.5cm},clip]{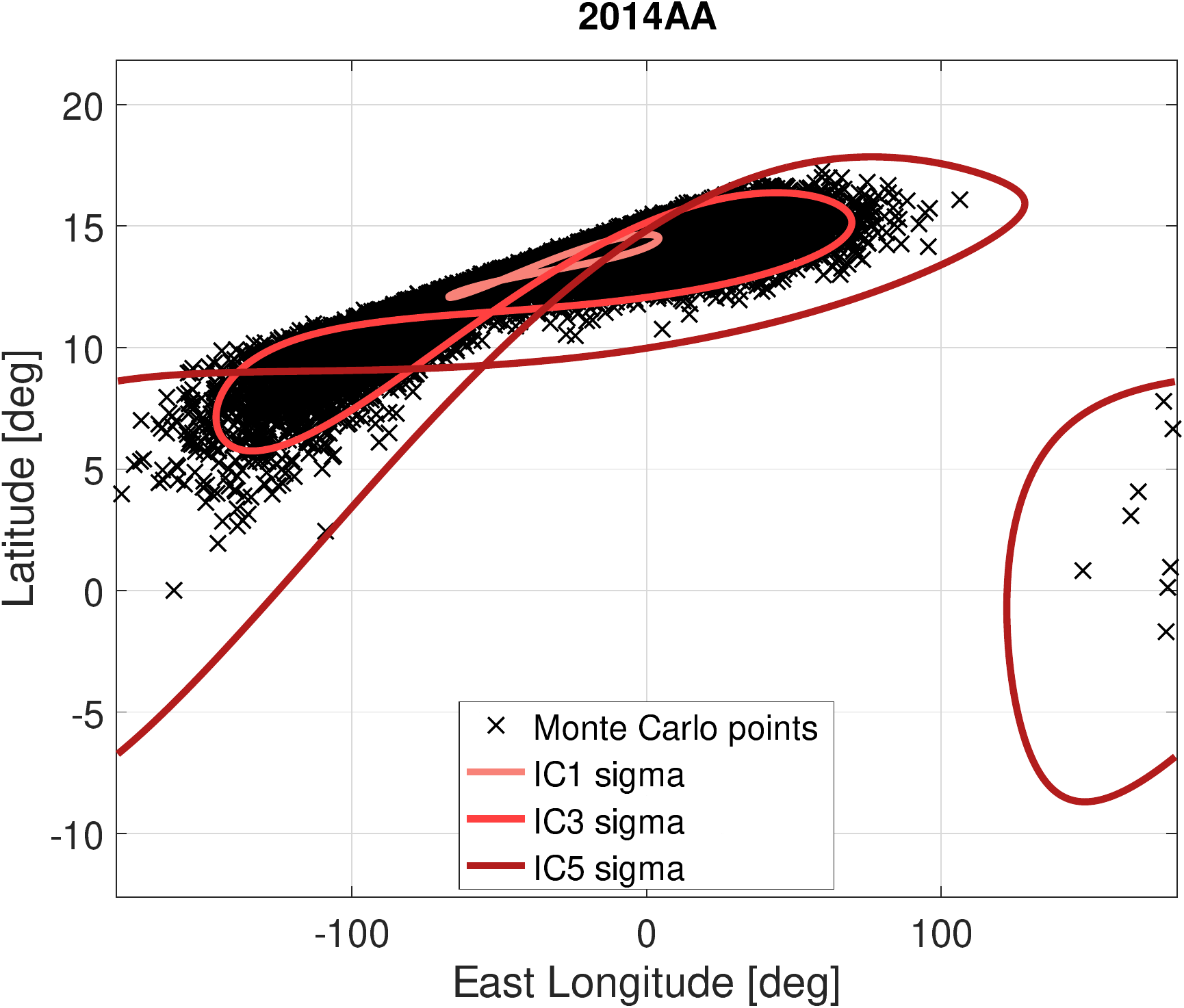}
     \caption{Impact region on ground of asteroid 2014~AA computed
       with the MOV method and by the semilinear method. {\em Left.}
       Comparison between the impact region computed with the MOV
       orbits having $\chi<5$ and the semilinear boundaries
       corresponding to the confidence levels $1$, $3$, and $5$. {\em
         Right.} Comparison between a Monte Carlo run and the same
       semilinear boundaries.}
    \label{fig:2014AA}
\end{figure}

Also in this case we can fit a nominal orbit and thus the MOV method
starts with the cobweb sampling of the AR. All the MOV orbits are
impacting, resulting in an impact probability of 100\%. The impact
regions on ground of 2014~AA computed with the MOV or with the Monte
Carlo method are shown in Figure~\ref{fig:2014AA}. Again we also show
the semilinear boundaries for comparison, this time corresponding to
the confidence levels $\sigma=1$, $3$, and $5$.

\noindent The plot deserves some comment to avoid misunderstadings,
due to the particular shape of the impact region. It is known that
semilinear boundaries are not necessarily simple curves. As a
consequence, if this is the case, the curve cannot be the boundary of
the non-linear prediction and indeed the impact regions extend outside
the drawn boundaries. This behaviour is confirmed by the fact that, in
the vicinity of the torsion, the boundaries with lower $\sigma$
extends outside the ones with higher $\sigma$. With a heuristic
approach we can state that the impact region extends outside the
region delimited by the boundary, on the side of the lower confidence
level boundaries. As a consequence this result is somewhat
unsatisfactory, because we can just guess the actual impact area. Of
course this issue disappears as soon as we start with a sampling of
the whole confidence region and not with just the sampling of a
curve. As we can see, the impact region computed with the MOV gives a
clear representation of the impact area and also shows why the
semilinear boundaries are twisted. Indeed the confidence region folds
on itself before being projected on the impact surface $S_h$, and this
is a consequence of the non-linear effects due to the ongoing close
approach. This behaviour is also confiermed by the Monte Carlo
simulation shown in the right plot of Figure~\ref{fig:2014AA}.

\subsection{Asteroid 2018~LA}
\label{sec:2018LA}

Asteroid 2018~LA was discovered by the Mt. Lemmon Observatory of the
Catalina Sky Survey just 8 hours before its impact in Botswana. For
this asteroid we also have the fireball report shown in
Table~\ref{tab:fireball}, which took place at 28.7~km of altitude.

\begin{figure}[h]
    \centering\includegraphics[scale=0.55,trim={0 0 0
      0.6cm},clip]{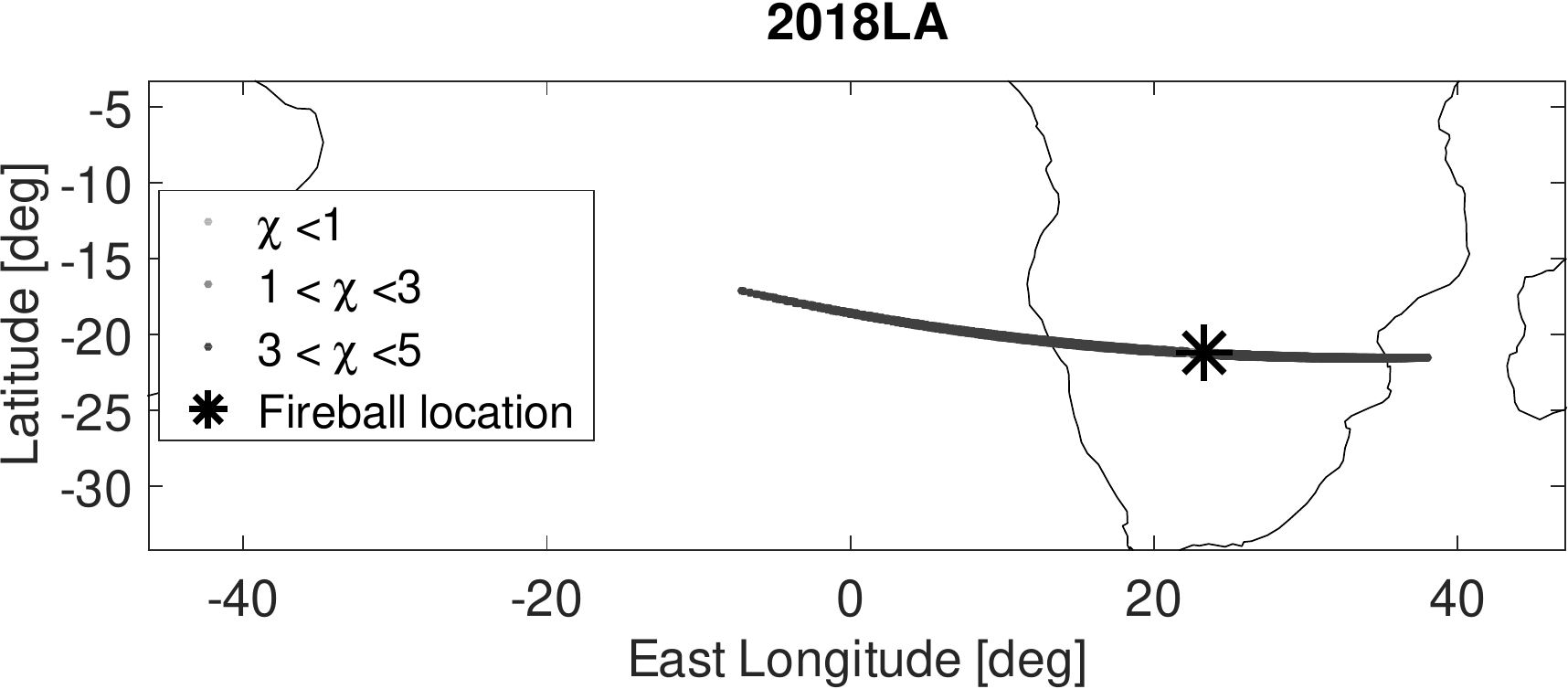}
    \caption{Impact region at $h=28.7$~km of 2018~LA computed with the
      MOV method starting from all the $14$ observations. The black
      star marks the location of the fireball event.}
    \label{fig:2018LA-1}
\end{figure}
\begin{figure}[h!]
    \includegraphics[scale=0.6,trim={0 0 0 0.5cm},clip]{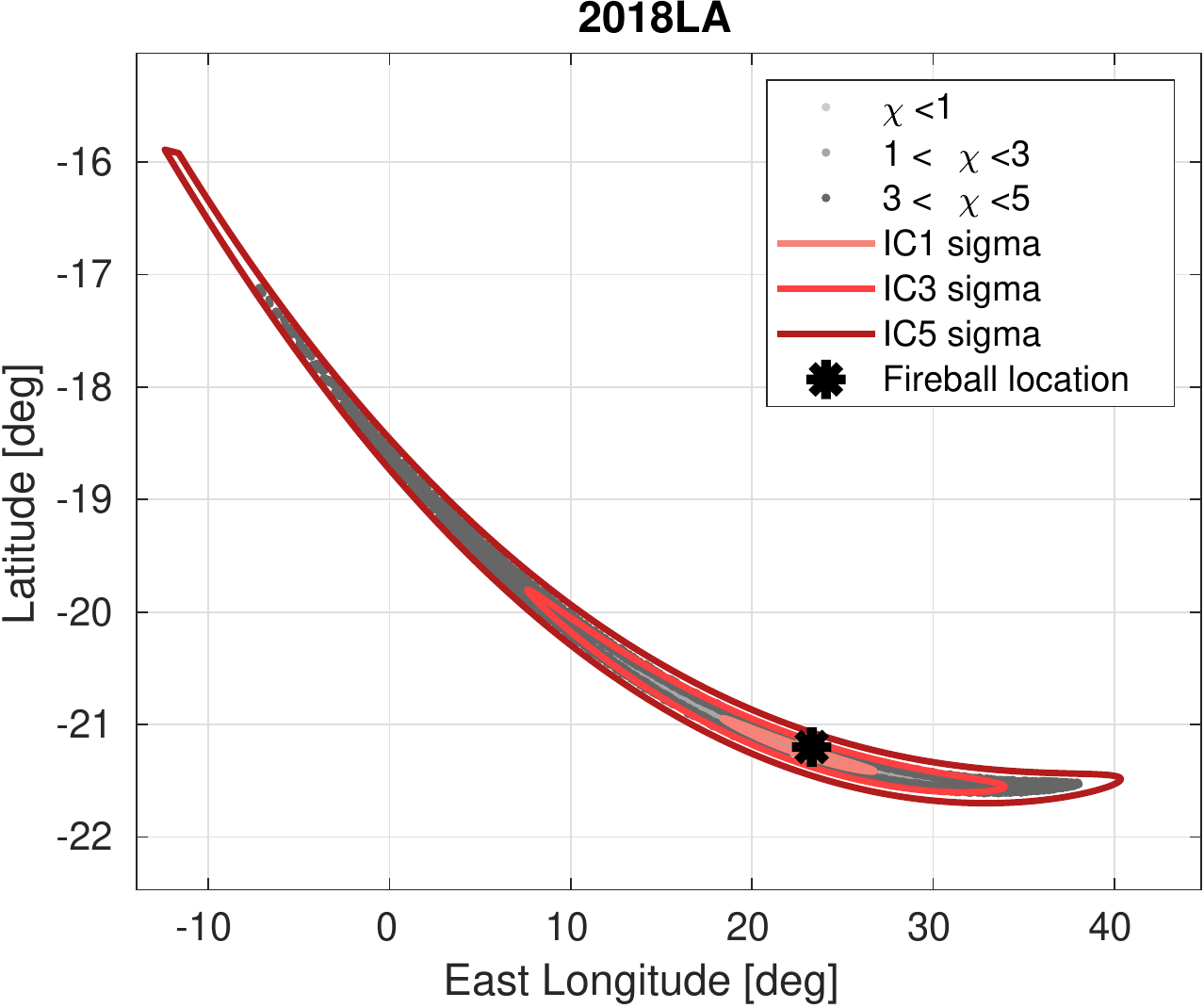}
    \hfill
    \includegraphics[scale=0.6,trim={0 0 0 0.5cm},clip]{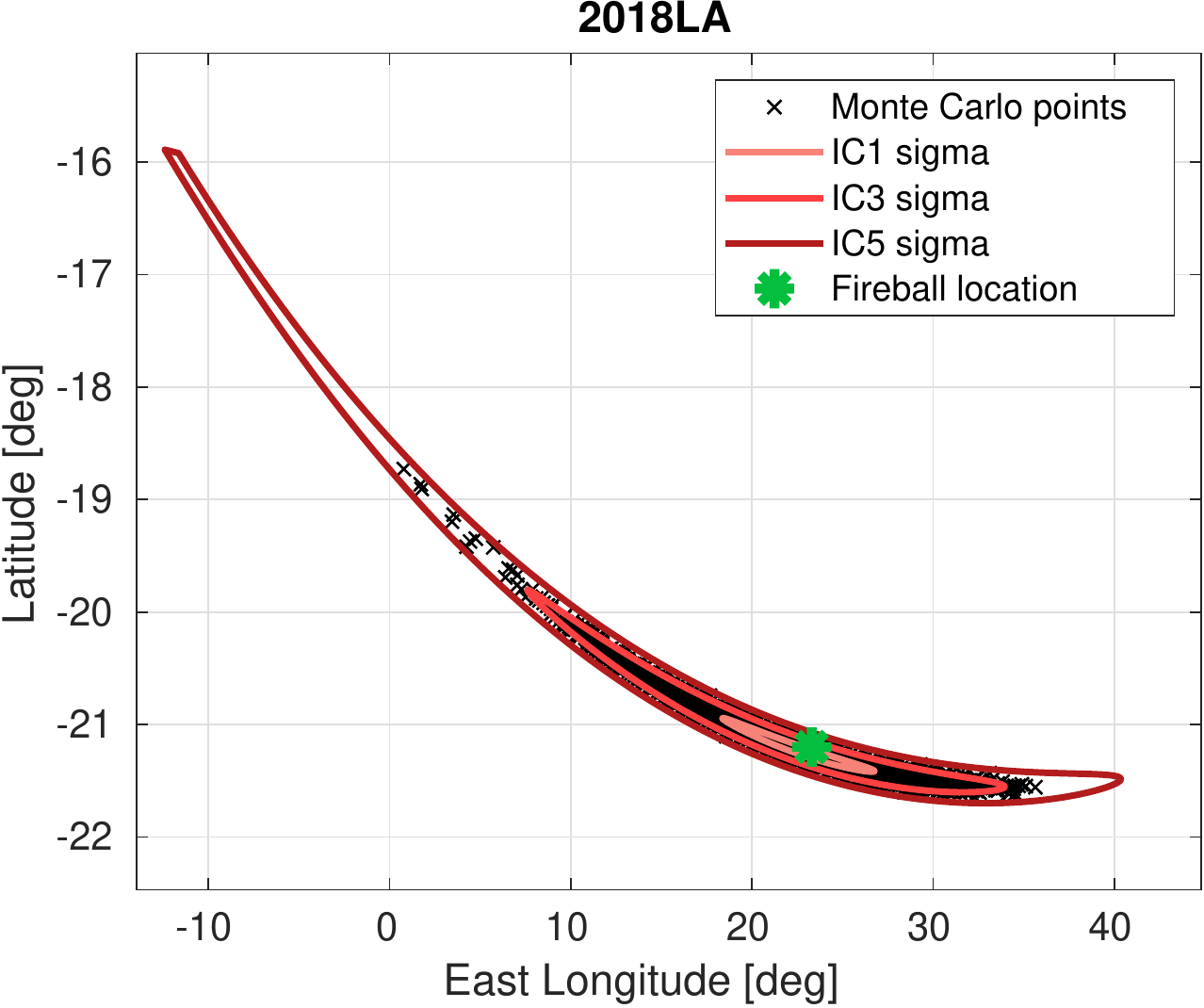}
    \caption{Impact location prediction at $h=28.7$~km for asteroid
      2018~LA, with all the $14$ observations. {\em Left.}  Comparison
      between the impact region computed with the MOV orbits having
      $\chi<5$ and the semilinear boundaries corresponding to the
      confidence levels $1$, $3$, and $5$. {\em Right.} Comparison
      between a Monte Carlo run and the same semilinear
      boundaries. The star marks the location of the fireball event.}
    \label{fig:2018LA-2}
\end{figure}

\noindent The astrometric dataset contains 14 observations, which are
enough to compute a full orbit and to apply the cobweb sampling. The
impact probability computed with the MOV method gives the certainty of
the impact and all the MOV orbits impact the
Earth. Figure~\ref{fig:2018LA-1} shows the MOV impact region at
$h=28.7$~km, together with the fireball
event. Figure~\ref{fig:2018LA-2} shows the comparison between the MOV
impact region or the Monte Carlo impact region against the semilinear
buondaries computed at the height of the fireball event. Also in this
case we have full agreement among the methods.

\begin{figure}[h!]
    \centering\includegraphics[scale=0.5,trim={0 0 0
      0.57cm},clip]{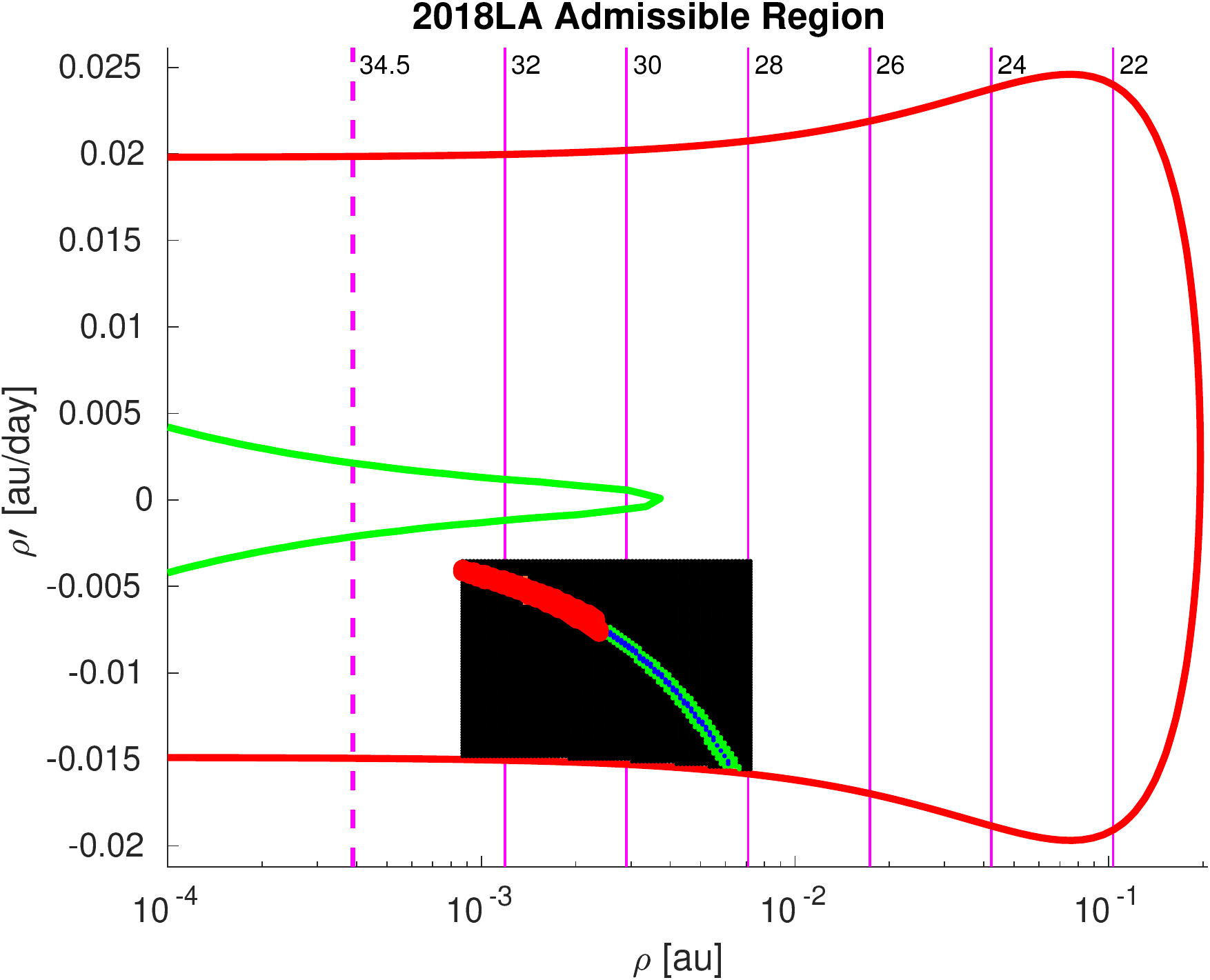}
    \caption{Second step of the Admissible Region grid sampling for
      asteroid 2018~LA with 12 observations.  The sample points are
      marked in blue when $\chi\leq 2$ and in green when $2<\chi<5$
      (note that the blue points are contained in the narrow region
      delimited by the green points). The red circles mark the
      impacting orbits.}
    \label{fig:2018LA-3a}
\end{figure}

\begin{figure}[h!]
    \centering\includegraphics[scale=0.75,trim={0 0 0 0.5cm},clip]{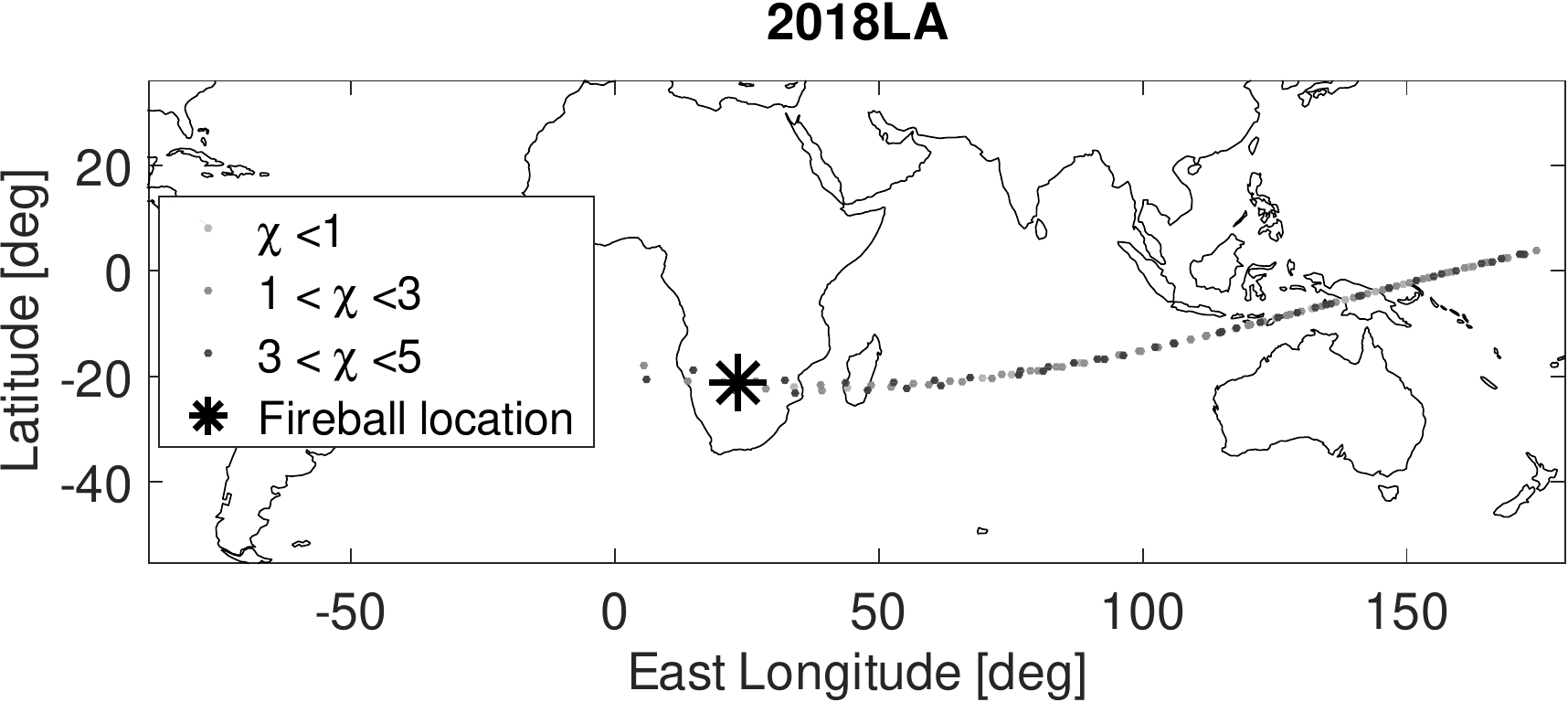}
    \caption{Impact region at altitude $h=28.7$~km of asteroid 2018~LA
      computed with the MOV method and using the first $12$
      observations. The star marks the location of the fireball event.}
    \label{fig:2018LA-3b}
\end{figure}

As pointed out in the introduction, the MOV method can be applied for
impact location predictions also when the amount of information
provided by the observational arc is not enough to constrain a full
orbit. In this case the semilinear method of \cite{dimare:imp_corr}
cannot be used, because it requires a nominal orbit and a virtual
impactor representative, as provided through the LOV method. To show
an example of such computation, we applied the MOV method to 2018~LA
with its first 12 observations. In this case the observations are not
enough to compute a reliable nominal orbit, thus the AR is sampled
with rectangular grids. The second grid samples a subregion of the AR,
with the orbits having $\chi<5$ spanning a quite narrow strip (see
Figure~\ref{fig:2018LA-3a}). The propagation of the impacting MOV
orbits gives the result shown in Figure~\ref{fig:2018LA-3b}, with the
impact region being quite elongated due to the poor constraint on the
asteroid orbit.

\subsection{Asteroid 2019~MO}
Asteroid 2019~MO was discovered from the ATLAS Mauna Loa observatory
on June 22, 2019, less than 12 hours before impacting the Earth
between Jamaica and the south American coast. Also in this case we
have a fireball event reported in Table~\ref{tab:fireball}, which took
place at 25~km of altitude.

\begin{figure}[h]
    \includegraphics[scale=0.9,trim={0 0 0 0.5cm},clip]{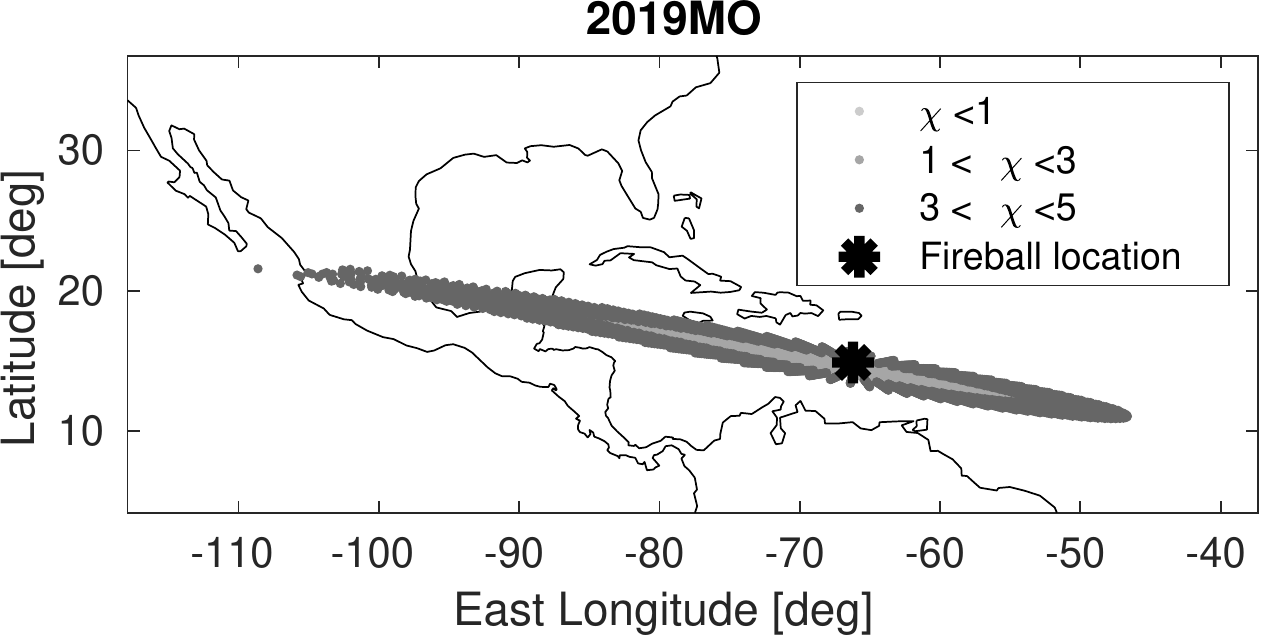}
    \caption{Impact region at altitude $h=25$~km of asteroid
      2019~MO. The black star marks the location of the fireball
      event.}
    \label{fig:2019MO-1}
\end{figure}

\begin{figure}[h]
    \includegraphics[scale=0.6,trim={0 0 0 0.5cm},clip]{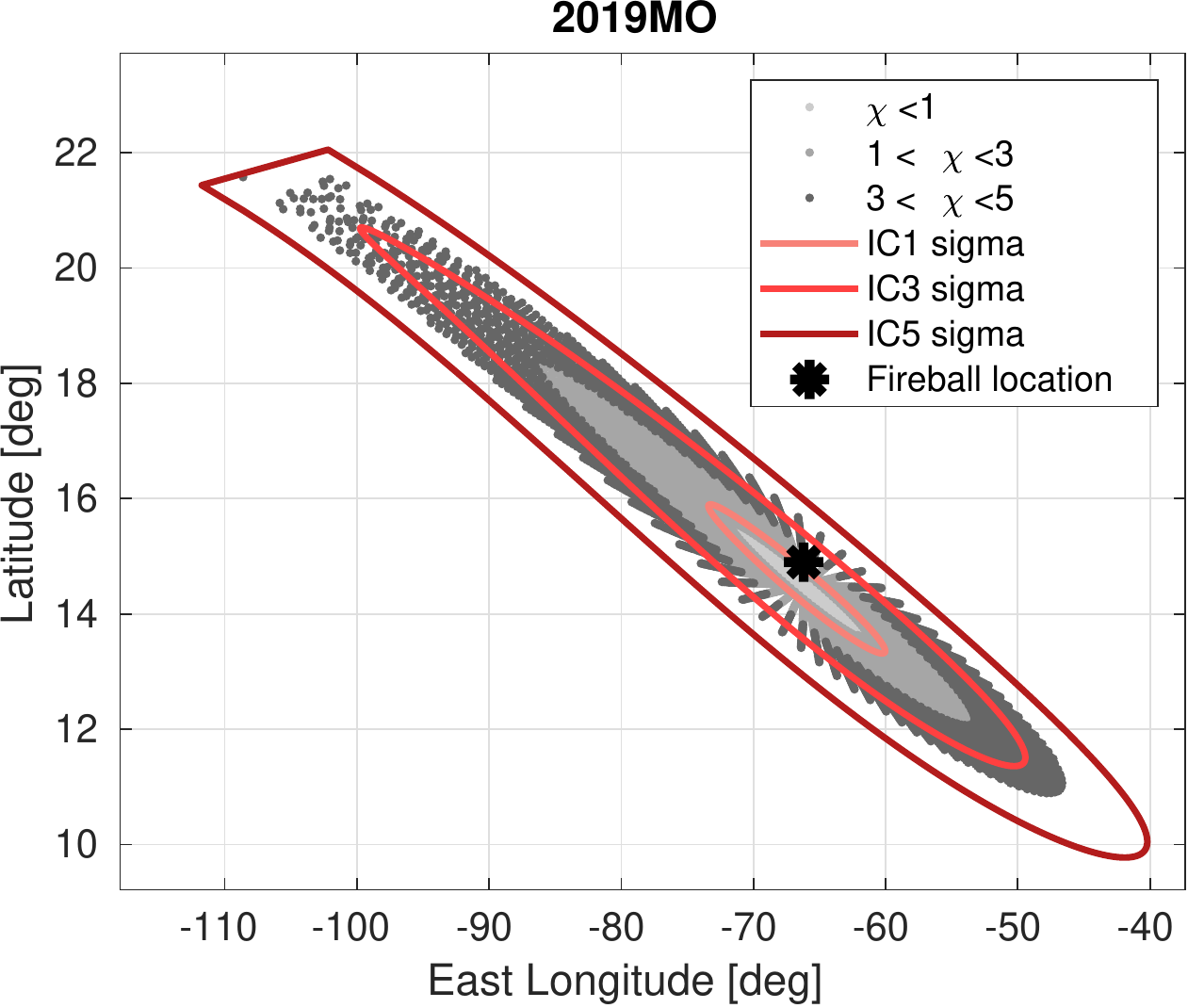}
    \hfill
    \includegraphics[scale=0.6,trim={0 0 0 0.5cm},clip]{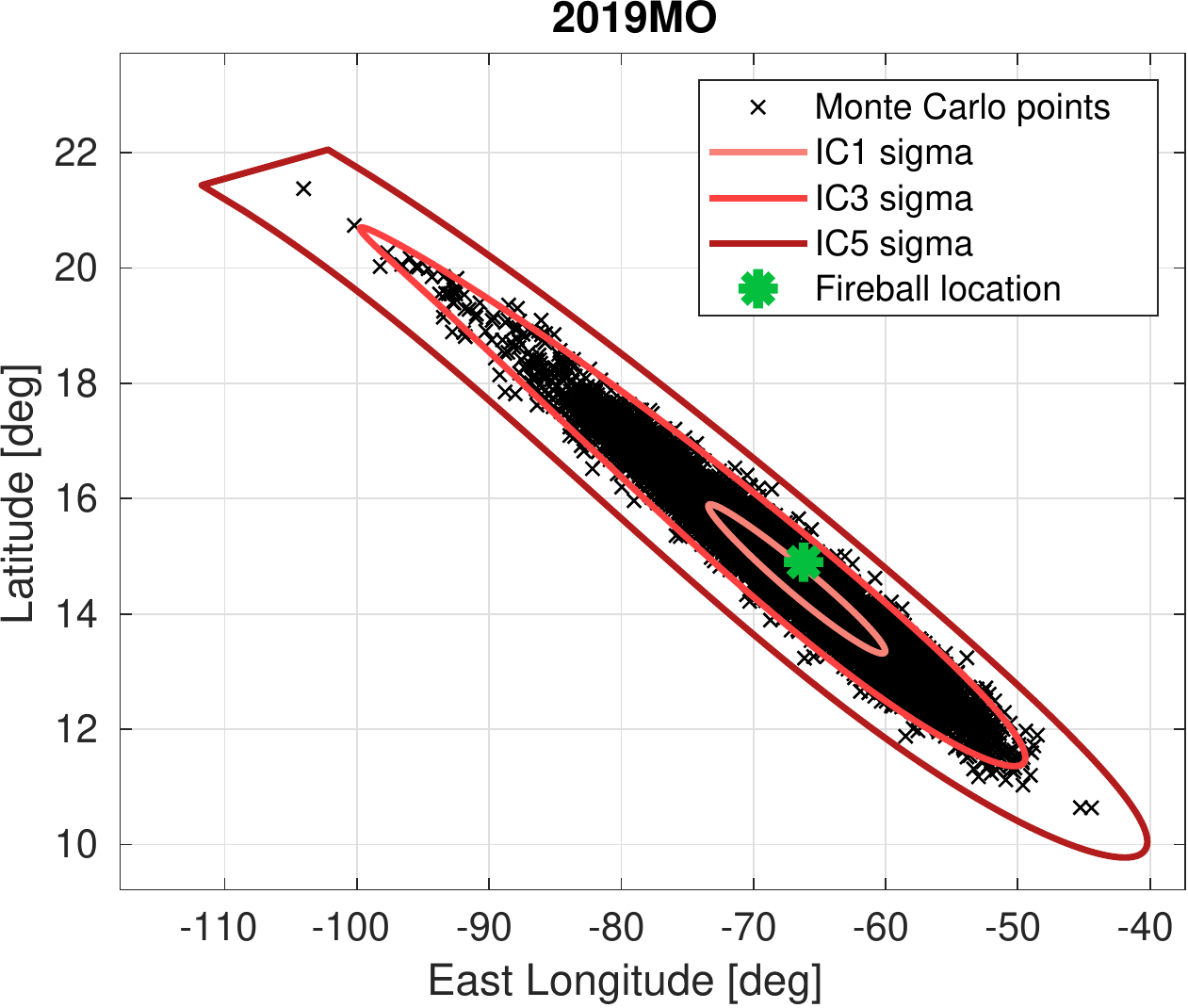}
    \caption{Impact location prediction at altitude $h=25$~km for
      asteroid 2019~MO. {\em Left.}  Comparison between the impact
      region computed with the MOV orbits having $\chi<5$ and the
      semilinear boundaries corresponding to the confidence levels
      $1$, $3$, and $5$. {\em Right.} Comparison between a Monte Carlo
      run and the same semilinear boundaries. The star marks the
      location of the fireball event.}
    \label{fig:2019MO-2}
\end{figure}

Figure~\ref{fig:2019MO-1} shows the MOV impact region at $h=25$~km
extending above the south American coast, together with the fireball
location. Figure~\ref{fig:2019MO-2} shows the comparison between the
MOV method, the semilinear boundaries and a Monte Carlo run with
10,000 sample points, which are fully compatible.

\begin{table}[!ht]
    \caption{Fireball reports corresponding to 2018~LA and 2019~MO
      impacts, extracted from the JPL web page
      \url{https://cneos.jpl.nasa.gov/fireballs/}.}
    \label{tab:fireball}
    \centering
    \small
    \begin{tabular}{ccccc}
        \toprule \textbf{Name} & \textbf{Peak Brightness Date/Time} &
        \textbf{Latitude} & \textbf{Longitude} &
        \textbf{Altitude}\\
        &(UTC) & (deg) & (deg) & (km)\\
        \midrule
        2018LA & 2018-06-02 16:44:12	& 21.2S	& 23.3E	& 28.7 \\
        \midrule
        2019MO & 2019-06-22 21:25:48	& 14.9N	& 66.2W & 25.0 \\
        \bottomrule
    \end{tabular}
\end{table}

\section{Conclusions}
\label{sec:conc}

Very small asteroids can be only observed during a deep close approach
with the Earth and it may be the case that an impact occurs a few days
after the discovery. In this paper we considered the problem of
predicting the impact location of such objects by exploiting the MOV,
a set of virtual asteroids representing the orbital uncertainty. Once
a MOV sampling is available it suffices to propagate each orbit for a
given amount of time and, for the impacting orbits, to compute the
geodetic coordinates of the impacting points. The impact region is
thus given by a set of points on the impact surface at a certain
height over the Earth.

The advantage of the MOV method with respect to already existing
techniques, like for example the semilinear method of
\cite{dimare:imp_corr}, is that it can be used even when it is
impossible to fit a nominal orbit to the few available astrometric
observations (see the example presented in Section~\ref{sec:2018LA}
and shown in Figure~\ref{fig:2018LA-3b}).

We tested our method using the data available for the four impacted
asteroids so far, namely 2008TC$_3$, 2014AA, 2018LA, and 2019MO. Since
these data are also enough to constrain a full orbit, we compared the
results of the MOV method with the semilinear boundaries and with the
outcome of an observational Monte Carlo simultation, getting a very
strong consistency among the three methods.


\bibliographystyle{elsarticle-harv} \bibliography{mov_impcor}

\begin{thebibliography}{16}
\expandafter\ifx\csname natexlab\endcsname\relax\def\natexlab#1{#1}\fi
\providecommand{\url}[1]{\texttt{#1}}
\providecommand{\href}[2]{#2}
\providecommand{\path}[1]{#1}
\providecommand{\DOIprefix}{doi:}
\providecommand{\ArXivprefix}{arXiv:}
\providecommand{\URLprefix}{URL: }
\providecommand{\Pubmedprefix}{pmid:}
\providecommand{\doi}[1]{\href{http://dx.doi.org/#1}{\path{#1}}}
\providecommand{\Pubmed}[1]{\href{pmid:#1}{\path{#1}}}
\providecommand{\bibinfo}[2]{#2}
\ifx\xfnm\relax \def\xfnm[#1]{\unskip,\space#1}\fi
\bibitem[{Del~Vigna(2020)}]{delvigna:mov1}
\bibinfo{author}{Del~Vigna, A.}, \bibinfo{year}{2020}.
\newblock \bibinfo{title}{{The Manifold Of Variations: hazard assessment of
  short-term impactors}}.
\newblock \bibinfo{journal}{Celestial Mechanics and Dynamical Astronomy}
  \bibinfo{volume}{132}.
\bibitem[{Del~Vigna et~al.(2020)Del~Vigna, Guerra and
  Valsecchi}]{delvigna:dens}
\bibinfo{author}{Del~Vigna, A.}, \bibinfo{author}{Guerra, F.},
  \bibinfo{author}{Valsecchi, G.B.}, \bibinfo{year}{2020}.
\newblock \bibinfo{title}{{Improving impact monitoring through LOV
  densification}}.
\newblock \bibinfo{journal}{Icarus} \bibinfo{volume}{351}.
\bibitem[{Del~Vigna et~al.(2019a)Del~Vigna, Milani, Spoto, Chessa and
  Valsecchi}]{delvigna:compl_IM}
\bibinfo{author}{Del~Vigna, A.}, \bibinfo{author}{Milani, A.},
  \bibinfo{author}{Spoto, F.}, \bibinfo{author}{Chessa, A.},
  \bibinfo{author}{Valsecchi, G.B.}, \bibinfo{year}{2019}a.
\newblock \bibinfo{title}{{Completeness of Impact Monitoring}}.
\newblock \bibinfo{journal}{Icarus} \bibinfo{volume}{{321}}.
\bibitem[{Del~Vigna et~al.(2019b)Del~Vigna, Roa, Farnocchia, Micheli, Tholen,
  Guerra, Spoto and Valsecchi}]{delvigna:410777}
\bibinfo{author}{Del~Vigna, A.}, \bibinfo{author}{Roa, J.},
  \bibinfo{author}{Farnocchia, D.}, \bibinfo{author}{Micheli, M.},
  \bibinfo{author}{Tholen, D.}, \bibinfo{author}{Guerra, F.},
  \bibinfo{author}{Spoto, F.}, \bibinfo{author}{Valsecchi, G.B.},
  \bibinfo{year}{2019}b.
\newblock \bibinfo{title}{{Yarkovsky effect detection and updated impact hazard
  assessment for near-Earth asteroid (410777)~2009~FD}}.
\newblock \bibinfo{journal}{Astronomy \& Astrophysics} \bibinfo{volume}{627},
  \bibinfo{pages}{A1}.
\bibitem[{Dimare et~al.(2020)Dimare, Del~Vigna, Bracali~Cioci and
  Bernardi}]{dimare:imp_corr}
\bibinfo{author}{Dimare, L.}, \bibinfo{author}{Del~Vigna, A.},
  \bibinfo{author}{Bracali~Cioci, D.}, \bibinfo{author}{Bernardi, F.},
  \bibinfo{year}{2020}.
\newblock \bibinfo{title}{{Use of a semilinear method to predict the impact
  corridor on ground}}.
\newblock \bibinfo{journal}{Celestial Mechanics and Dynamical Astronomy}
  \bibinfo{volume}{132}.
\bibitem[{{Farnocchia} et~al.(2015){Farnocchia}, {Chesley} and
  {Micheli}}]{farnocchia2015}
\bibinfo{author}{{Farnocchia}, D.}, \bibinfo{author}{{Chesley}, S.R.},
  \bibinfo{author}{{Micheli}, M.}, \bibinfo{year}{2015}.
\newblock \bibinfo{title}{{Systematic ranging and late warning asteroid
  impacts}}.
\newblock \bibinfo{journal}{Icarus} \bibinfo{volume}{258},
  \bibinfo{pages}{18--27}.
\bibitem[{Farnocchia et~al.(2017)Farnocchia, Jenniskens, Robertson, Chesley,
  Dimare and Chodas}]{farnocchia2008}
\bibinfo{author}{Farnocchia, D.}, \bibinfo{author}{Jenniskens, P.},
  \bibinfo{author}{Robertson, D.K.}, \bibinfo{author}{Chesley, S.R.},
  \bibinfo{author}{Dimare, L.}, \bibinfo{author}{Chodas, P.W.},
  \bibinfo{year}{2017}.
\newblock \bibinfo{title}{{The impact trajectory of asteroid 2008 TC$_3$}}.
\newblock \bibinfo{journal}{Icarus} \bibinfo{volume}{294},
  \bibinfo{pages}{218--226}.
\bibitem[{Milani et~al.(2005a)Milani, Chesley, Sansaturio, Tommei and
  Valsecchi}]{milani:clomon2}
\bibinfo{author}{Milani, A.}, \bibinfo{author}{Chesley, S.},
  \bibinfo{author}{Sansaturio, M.E.}, \bibinfo{author}{Tommei, G.},
  \bibinfo{author}{Valsecchi, G.B.}, \bibinfo{year}{2005}a.
\newblock \bibinfo{title}{{Nonlinear impact monitoring: line of variation
  searches for impactors}}.
\newblock \bibinfo{journal}{Icarus} \bibinfo{volume}{173},
  \bibinfo{pages}{362--384}.
\bibitem[{{Milani} et~al.(2004){Milani}, {Gronchi}, {De' Michieli Vitturi} and
  {Kne\^zevi\'c}}]{milani2004AR}
\bibinfo{author}{{Milani}, A.}, \bibinfo{author}{{Gronchi}, G.F.},
  \bibinfo{author}{{De' Michieli Vitturi}, M.},
  \bibinfo{author}{{Kne\^zevi\'c}, Z.}, \bibinfo{year}{2004}.
\newblock \bibinfo{title}{{Orbit determination with very short arcs. {I}
  admissible regions}}.
\newblock \bibinfo{journal}{Celestial Mechanics and Dynamical Astronomy}
  \bibinfo{volume}{90}, \bibinfo{pages}{57--85}.
\bibitem[{{Milani} et~al.(2007){Milani}, {Gronchi} and {Kne{\v
  z}evi{\'c}}}]{milani2007}
\bibinfo{author}{{Milani}, A.}, \bibinfo{author}{{Gronchi}, G.F.},
  \bibinfo{author}{{Kne{\v z}evi{\'c}}, Z.}, \bibinfo{year}{2007}.
\newblock \bibinfo{title}{{New Definition of Discovery for Solar System
  Objects}}.
\newblock \bibinfo{journal}{Earth Moon and Planets} \bibinfo{volume}{100},
  \bibinfo{pages}{83--116}.
\bibitem[{Milani et~al.(2005b)Milani, Sansaturio, Tommei, Arratia and
  Chesley}]{milani:multsol}
\bibinfo{author}{Milani, A.}, \bibinfo{author}{Sansaturio, M.},
  \bibinfo{author}{Tommei, G.}, \bibinfo{author}{Arratia, O.},
  \bibinfo{author}{Chesley, S.R.}, \bibinfo{year}{2005}b.
\newblock \bibinfo{title}{{Multiple solutions for asteroid orbits:
  {C}omputational procedure and applications}}.
\newblock \bibinfo{journal}{Astronomy \& Astrophysics} \bibinfo{volume}{431},
  \bibinfo{pages}{729--746}.
\bibitem[{{Milani} and {Valsecchi}(1999)}]{milani:ident2}
\bibinfo{author}{{Milani}, A.}, \bibinfo{author}{{Valsecchi}, G.B.},
  \bibinfo{year}{1999}.
\newblock \bibinfo{title}{{The Asteroid Identification Problem. II. Target
  Plane Confidence Boundaries}}.
\newblock \bibinfo{journal}{Icarus} \bibinfo{volume}{140},
  \bibinfo{pages}{408--423}.
\bibitem[{{NIMA - National Imagery and Mapping Agency}(2000)}]{wgs84}
\bibinfo{author}{{NIMA - National Imagery and Mapping Agency}},
  \bibinfo{year}{2000}.
\newblock \bibinfo{title}{{Technical Report 8350.2~. Third Edition. Department
  of Defence World Geodetic System 1984. Its Definition and Relationships with
  Local Geodetic Systems}}.
\newblock \bibinfo{type}{Technical Report}. {NIMA - National Imagery and
  Mapping Agency}.
\bibitem[{Spoto et~al.(2018)Spoto, Del~Vigna, Milani, Tommei, Tanga, Mignard,
  Carry, Thuillot and David}]{spoto:immimp}
\bibinfo{author}{Spoto, F.}, \bibinfo{author}{Del~Vigna, A.},
  \bibinfo{author}{Milani, A.}, \bibinfo{author}{Tommei, G.},
  \bibinfo{author}{Tanga, P.}, \bibinfo{author}{Mignard, F.},
  \bibinfo{author}{Carry, B.}, \bibinfo{author}{Thuillot, W.},
  \bibinfo{author}{David, P.}, \bibinfo{year}{2018}.
\newblock \bibinfo{title}{{Short arc orbit determination and imminent impactors
  in the Gaia era}}.
\newblock \bibinfo{journal}{Astronomy \& Astrophysics} \bibinfo{volume}{{614}}.
\bibitem[{Tholen and Farnocchia(2020)}]{tholen:99942}
\bibinfo{author}{Tholen, D.}, \bibinfo{author}{Farnocchia, D.},
  \bibinfo{year}{2020}.
\newblock \bibinfo{title}{{Detection of Yarkovsky Acceleration of (99942)
  Apophis}}.
\newblock \bibinfo{journal}{Bulletin of the AAS} \bibinfo{volume}{52}.
\bibitem[{Tommei(2006)}]{tommei:phd}
\bibinfo{author}{Tommei, G.}, \bibinfo{year}{2006}.
\newblock \bibinfo{title}{{Impact Monitoring of NEOs: theoretical and
  computational results}}.
\newblock Ph.D. thesis. University of Pisa.

\end{thebibliography}

\end{document}